# Improving image quality of the Solar Disk Imager (SDI) of the Lyman-alpha Solar Telescope (LST) onboard the ASO-S mission

Hui Liu[1] ⊙, Hui Li[2,3] ⊙, Sizhong Zou[1] ⊙, Kaifan Ji[1] ⊙, Zhenyu Jin[1] ⊙, Jiahui Shan[2] ⊙, Jingwei Li[2] ⊙, Guanglu Shi[2] ⊙, Yu Huang[2,3] ⊙, Li Feng[2,3] ⊙, Jianchao Xue[2] ⊙, Qiao Li[2] ⊙, Dechao Song[2] ⊙ and Ying Li[2] ⊙

[1] Yunnan Observatories, Chinese Academy of Sciences, Kunming 650216, China;

[2] Key Laboratory of Dark Matter and Space Astronomy, Purple Mountain Observatory, Chinese Academy of Sciences, Nanjing 210023, China; *nj.lihui@pmo.ac.cn*

[3] School of Astronomy and Space Science, University of Science and Technology of China, Hefei 230026, China

Received 20xx month day; accepted 20xx month day

Abstract The in-flight calibration and performance of the Solar Disk Imager (SDI), which is a pivotal instrument of the Lyman-alpha Solar Telescope (LST) onboard the Advanced Space-based Solar Observatory (ASO-S) mission, suggested a much lower spatial resolution than expected. In this paper, we developed the SDI point-spread function (PSF) and Image Bivariate Optimization Algorithm (SPIBOA) to improve the quality of SDI images. The bivariate optimization method smartly combines deep learning with optical system modeling. Despite the lack of information about the real image taken by SDI and the optical system function, this algorithm effectively estimates the PSF of the SDI imaging system directly from a large sample of observational data. We use the estimated PSF to conduct deconvolution correction to observed SDI images, and the resulting images show that the spatial resolution after correction has increased by a factor of more than three with respect to the observed ones. Meanwhile, our method also significantly reduces the inherent noise in the observed SDI images. The SPIBOA has now been successfully integrated into the routine SDI data processing, providing important support for the scientific studies based on the data. The development and application of SPIBOA also pave new ways to identify astronomical telescope systems and enhance observational image quality. Some essential factors and precautions in applying the SPIBOA method are also discussed.

## 1 INTRODUCTION

The Advanced Space-based Solar Observatory (ASO-S; Gan et al. 2019, 2023), launched on 2022 October 9, is the first comprehensive Chinese space observatory dedicated to observations of the Sun. The Lymanalpha Solar Telescope (LST), characterized by its unique observations in the Hydrogen Lyman-alpha line (121.6 nm), the most intensive line in ultraviolet (UV) band of solar spectrum, from disk center to inner corona up to 2.5 solar radii, is one of the three payloads onboard ASO-S mission and was described in detail in Li et al. (2019) and Chen et al. (2019). Here, we provide brief introduction to the LST payload, which is essential for readers to understand the study in this paper.

The LST payload comprises three instruments:

- **White-light Solar Telescope (WST)**: With an aperture of 130 mm, the WST works in the 360.0±1.9 nm waveband to observe white-light flares on the Sun with a field of view (FOV) up to 1.2 solar radii. The highest cadence of WST is 1 – 2 seconds in burst mode.

- **Dual-channel Solar Corona Imager (SCI):** Featuring an aperture of 60 mm, the SCI works in both the Hydrogen Lyman-alpha (SCIUV: 122.2±3.9 nm) and the visible (SCIWL: 704.1±31.3 nm) wavebands to captures coronal mass ejections (CMEs) and other activities within the inner corona up to 2.5 solar radii.

- **Solar Disk Imager (SDI)**: The SDI has an aperture of 68 mm and works in the Lyman-alpha waveband similar to SCIUV but with slightly different central wavelength and width (120.8±4.5 nm). It is used to observe various activities on the solar disk with the same FOV as the WST.

The in-flight calibration and performance of the LST have been successfully completed and presented in Chen et al. (2024). Readers are referred to it for detailed information.

SDI is proposed to image the Sun with a spatial resolution of $1.2''$ and a cadence of as high as 4 seconds in burst mode and a lower cadence (say 60 seconds) in routine mode depending on the available telemetry (Li et al. 2019; Chen et al. 2019). As outlined in Chen et al. (2024), in-flight observations revealed a much lower spatial resolution than expected, which is measured to be approximately $9.5''$ based on first-light images captured on 2022 October 26. One potential cause suggested is wavefront errors in the entrance Lyman-alpha filter, which could lead to defocusing and subsequent image blurring. However, it's important to note that other factors can also contribute to such a low measured spatial resolution, including, but not limited to, aberrations imperfections in the optical components of the instrument. To determine the exact causes of the lower resolution is beyond the scope of this paper.

SDI images sometimes exhibit horseshoe-like bright structures, as illustrated in the white rectangular area of Figure 1. These structures, together with the low spatial resolution, significantly affect research efforts related to the scientific objectives of LST, including the evolution and eruption of filaments/prominences, the characteristics of flares and the source regions of CMEs in the Lyman-alpha line, the parameters and triggering mechanisms of solar flares, and potential wave phenomena in the Lymanalpha line on the solar disk (Li et al. 2019; Feng et al. 2019). Therefore, it is crucial to mitigate or eliminate these structures and enhance image quality (in terms of spatial resolution, noise reduction, etc.) through various algorithms. A comparison of SDI images with those captured by the Atmospheric Imaging Assembly

(AIA, Lemen et al. 2012) on the Solar Dynamics Observatory (SDO, Pesnell et al. 2012) in the 304 Å and˚

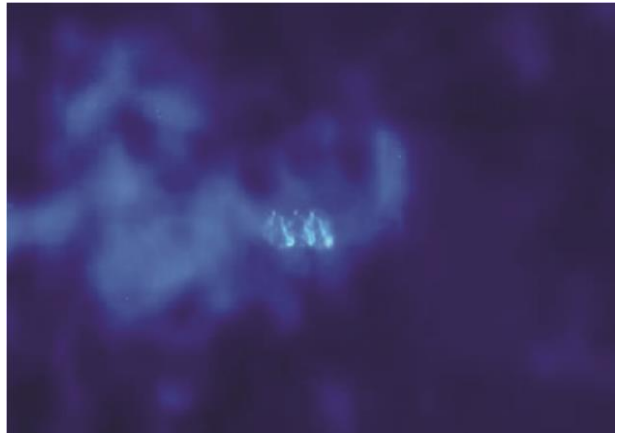

Fig.1: A cutout of SDI sample image showing 'horseshoe'brightenings.

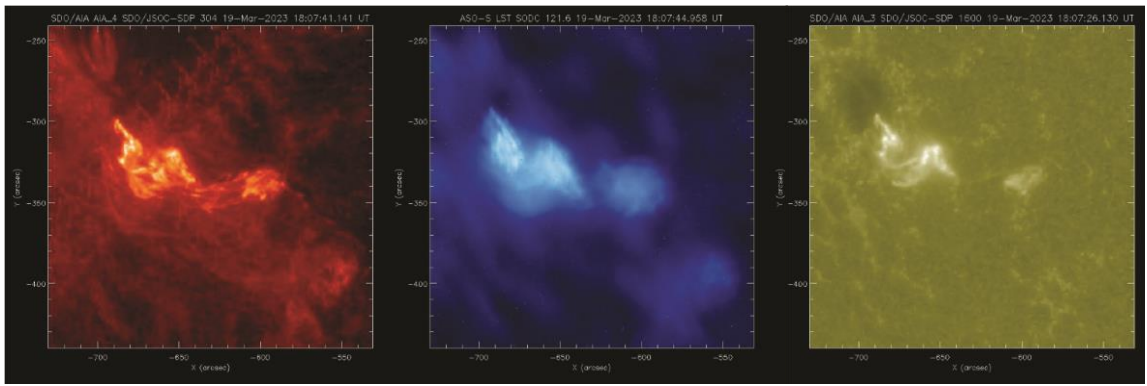

Fig.2: Comparison of AIA 304 A(left), SDI (middle) and AIA 1600° A (right) images at 18:07 UT on 2023° March 19.

1600 A channels, as shown in Figure 2, reveals that the sharp, bright structures in the AIA images appear° diffused and blurred in the SDI images, further emphasizing the necessity to improve the quality of SDI images.

One promising algorithm to enhance image quality relies on deep learning techniques (LeCun et al. 2015), which have experienced rapid advancements in recent years and had widespread applications in astronomical adaptive optics. In this paper, we leverage deep learning techniques and in-flight observational data, taking into account the optical design of the SDI, to estimate the wavefront aberration of the SDI imaging system, and subsequently apply the results to observational data to improve image quality. The reconstructed images exhibit significant quality improvement compared to the original ones, demonstrating the feasibility of our proposed method.

This paper is organized as follows. Section 2 introduces the development and application of adaptive optics and wavefront sensing technologies based on deep learning. We describe our basic methodology and estimate the point-spread function (PSF) of SDI imaging system in Section 3, and delineate our experiment and analysis in Section 4. In Sections 5 and 6, we present our discussions and conclusions, respectively.

## 2  DEEP LEARNING FOR ADAPTIVE OPTICS AND WAVEFRONT SENSING

As a core technology of artificial intelligence, deep learning has significant advances in many fields, including computer vision, natural language processing, bioinformatics and drug discovery (LeCun et al. 2015). Deep learning, powered by artificial neural networks, is accomplished in processing complicate data through its ability to learn sophisticated multi-level representations and abstractions. These multiple layers of nonlinear transformations, also known as deep neural networks (DNNs), enable the extraction of intricate patterns from data.

Improving the accuracy and sensitivity of wavefront sensors in adaptive optics systems through deep learning not only represents a developmental trend in the field but also may establish a novel domain within its applications. This is to enhance wavefront detection accuracy and expand the capability of wavefront sensors to handle complex scenarios.

Deep learning technology was first applied to astronomical adaptive optics in the early 1990s, marking a significant development in the field. Angel et al. (1990) successfully used artificial neural networks to correct for piston and tip/tilt aberrations in the Multiple Mirror Telescope (MMT), mitigating the effects of turbulence and tilt between mirrors. Sandler et al. (1991) further applied neural network wavefront detection on a 1.5 m single-mirror telescope at the Phillips Laboratory of the United States Air Force. This technique was later extended to the Hubble Space Telescope (Barrett & Sandler 1993) for estimating static aberrations, demonstrating its adaptability to complex applications and its emergence in a novel application domain.

Recent advancements in deep learning adaptive optics have led to significant breakthroughs, improving observation accuracy, extending range, enhancing cost-effectiveness, and broadening applicability, ultimately strengthening astronomical observations. Notably, Nishizaki et al. (2019) introduced a machine learning based wavefront sensor that streamlines optical hardware and image processing, demonstrating the ability to estimate Zernike coefficients directly from single images using convolutional neural networks (CNNs). They also demonstrated that the sensor could be trained to estimate wavefront distortions from both point and extended sources.

Niu et al. (2020) explored deep learning algorithms for interferometric wavefront detection, accurately extracting phase distribution and analyzing distortions under general conditions. Their method has been experimentally validated for higher measurement accuracy, faster computation, and excellent performance in noisy conditions. A deep learning based method was proposed by Wang et al. (2021) for wavefront sensing and aberration correction in atmospheric turbulence. They successfully recovered the wavefront distortion phase from distorted images and corrected them using spatial light modulators, significantly improving image quality and resolution, providing a new solution for adaptive optical systems. Furthermore, de Bruijne et al. (2022) investigated deep learning wavefront sensing for extended sources, integrating blind deconvolution with deep learning to process Shack-Hartmann sensor images, which enhances precision and reliability in complicate application scenarios and wavefront distortions. Lastly, Ge et al. (2023) presented a targetindependent dynamic wavefront sensing method, which employed a distorted grating and deep learning to enable real-time detection and correction of aberrations across various and dynamic environments.

Recent researches in wavefront sensing using deep learning, as mentioned, have yielded significant achievements and wide-range practical applications. These studies not only introduce innovative methods and technologies but confirm their practicality and efficacy through experimental validation as well. These findings are pivotal for advancing astronomical adaptive optics and wavefront sensing technologies. Moreover, they offer invaluable insights into our research to enhance the quality of SDI observational images through wavefront aberration correction.

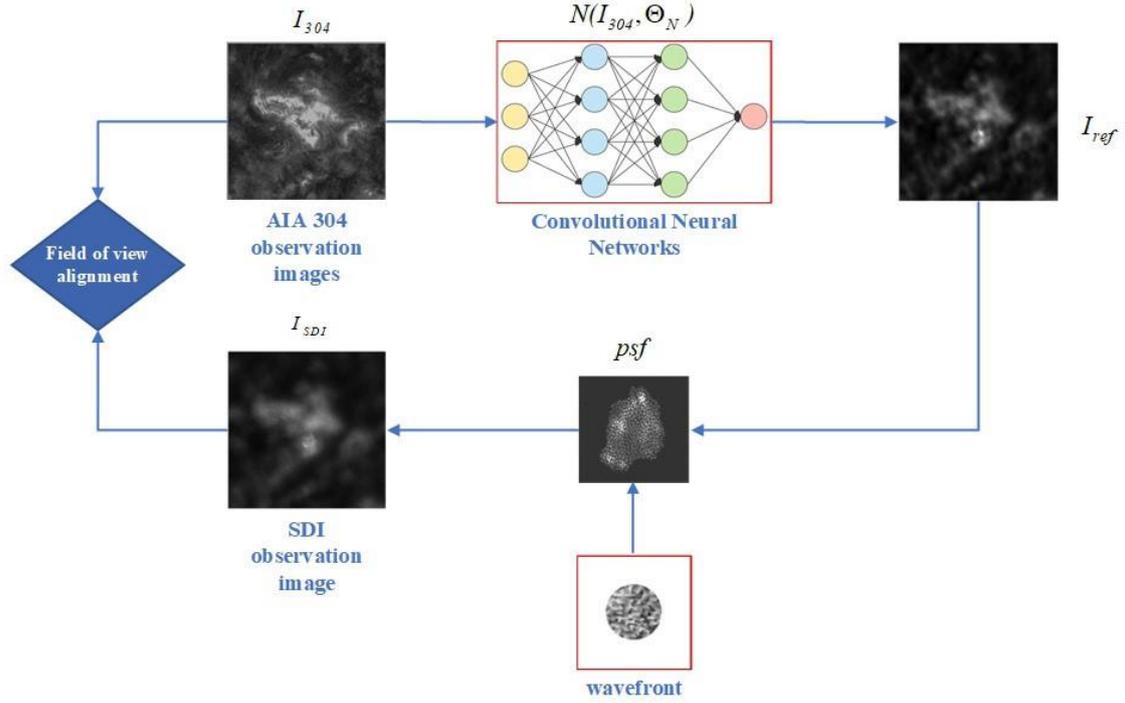

Fig.3: Flow chart of instrumental *psf* estimating process.

## 3 INSTRUMENTAL POINT SPREAD FUNCTION ESTIMATE

This section outlines the fundamental approach and method for applying deep learning techniques in our study, along with the process for estimating the instrumental point spread function (*psf*).

To mitigate the risk of unpredictable nonlinear distortions, we chose to keep the SDI optical imaging system within the linear convolution domain. By employing a structured approach, we apply a bivariate optimization strategy to simultaneously estimate both the true SDI image $I_{ref}$ and the *psf*, which are unknowns in our calculations. We translated the *psf* estimation into a wavefront estimation of the imaging system so as to effectively leverage the optical parameter information from the SDI imaging system.

Unlike the deep learning based methods for imaging system estimation reviewed in Section 2, we do not use CNN for direct wavefront prediction. Instead, we utilize CNN to apply nonlinear transformations to high-quality observational images with similar characteristics, such as those from AIA 304 A channel, to˚ generate ideal SDI reference images that comply with the linear constraints of the optical system. The CNN used in this study denoted with $N(I_{304}, \Theta_N)$ is a standard multi-layer CNN with an input layer consisting of 1 to 64 convolutional kernels, which transform the input image $I_{304}$ into 64 feature maps. $N(I_{304}, \Theta_N)$ contains 72 hidden layers with a width of 64. The output layer uses 64 to 1 convolutional kernel to transform the 64 feature maps into the output image $I_{ref}$. The algorithmic process for estimating the *psf* of the SDI imaging system is illustrated in Figure 3.

The objective of the optimization is to iteratively refine the parameter $\Theta_N$ of the convolutional neural network $N(I_{304}, \Theta_N)$ and the value of the *wavefront* to minimize the error defined by Equation (1), ensuring that the convolution of the ideal SDI image $I_{ref}$ with the system's *psf* aligns with the actual observational data $I_{SDI}$, thereby enabling an accurate estimation of the *psf*.

$$e = \min_{\Theta_N, wavefront} ||I_{SDI} - I_{ref} \otimes psf||^2 \tag{1}$$

The ideal SDI image $I_{ref}$ is output by the neural network $N(I_{304}, \Theta_N)$ (Equation (2)).

$$I_{ref} = N(I_{304}, \Theta_N) \tag{2}$$

The system's *psf* is calculated from the generalized pupil function defined by Equation (3):

$$psf = \left| \mathcal{F}\left( A \times e^{j \times wavefront} \right) \right|^2 \tag{3}$$

The $\Theta_N$ in these equations denotes the parameters of the CNN, while *wavefront* refers to the wavefront aberration, identified as the variable to be optimized and highlighted in the red square in Figure 3. The term $A$ stands for the aperture transmission, which equals 1 inside the aperture and 0 outside, assuming uniform transmission. The function F() denotes the Fourier transform, while the symbol $\otimes$ signifies the convolution operator.

Figure 3 depicts the processing steps of the SPIBOA (SDI PSF and Image Bivariate Optimization Algorithm). The algorithm's detailed procedure is as follows:

1) Data Preprocessing: coalign the images of SDI and AIA, and suppress noise in the SDI observational images.

2) Initialize $I_{ref}$ with the AIA 304 A image˚ $I_{304}$ and the *wavefront* with the data gleaned from pre-launch (on-ground) tests of the SDI flight model. 3)

Calculate the error

$$e = ||_{SDI} - I_{ref} \otimes psf||^2 \tag{4}$$

4) By propagating the error $e$ backward, the stochastic optimization algorithm (Adam) algorithm (Kingma & Ba 2014) is utilized to update the optimization targets, namely $I_{ref}$ and *wavefront*.

5) If $e$ is less than the predefined threshold, terminate the iteration and output the *psf*. Otherwise, continue iterating and return to step 4).

SPIBOA accepts SDI and AIA 304 A observational images with aligned FOV and suppressed thermal˚ noise as input. It employs the Adam algorithm to jointly refine the wavefront aberration (*wavefront*) and the parameters of the CNN ($\Theta_N$). By minimizing the error defined in Equation (1), SPIBOA derives an estimate of the ideal SDI image ($I_{ref}$) and the wavefront aberration (*wavefront*). Subsequently, it computes the *psf* of the SDI observational system using Equation (3). This *psf* is then utilized for deconvolution of the observed SDI images to enhance their quality.

Figure 4 depicts the intermediary results of the SPIBOA training procedure. Panel (a) displays the $I_{SDI}$ from the training dataset, while panel (b) shows the result of the convolution between $I_{ref}$ and *psf*. Panel (c) presents the $I_{ref.}$ generated by the CNN $N(I_{304}, \Theta_N)$, and panel (d) gives the initial value of $I_{ref}$, namely $I_{304}$.

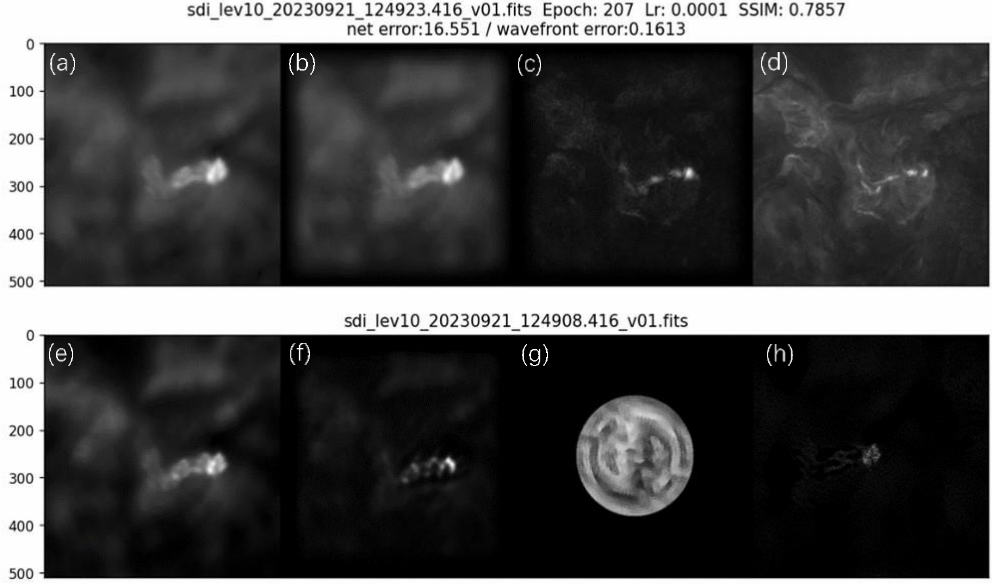

Fig.4: Intermediate Results of SPIBOA Training Process. (a) SDI observational image for training; (b) $I_{ref} \otimes psf$; (c) $I_{ref}$; (d) $I_{304}$; (e) SDI observational image for testing; (f) deconvolution of the test image with the obtained $psf$; (g) *wavefront* generated by SPIBOA; (h) *psf* obtained from the *wavefront* using Equation (3).

The resemblance between panels (b) and (a) suggests that the optimal solution derived by SPIBOA aligns with the SDI imaging model. Panels (e) and (f) present a test sample that was excluded from the training phase, together with its deconvolution outcomes, thereby implying the generalization capability of SPIBOA. Panel (g) presents the *wavefront* optimized by SPIBOA, wheras panel (h) displays the *psf* derived from this *wavefront* based on Equation (3). This *psf* is consistent with the anomalous 'horseshoe' structure depicted in Figure 1, suggesting that the estimation of the *psf* is reasonable.

## 4  EXPERIMENT AND ANALYSIS

We use the *psf* derived from SPIBOA to make deconvolution correction to actual SDI observational images utilizing the Richardson-Lucy iteration algorithm (Richardson 1972), and present the results in Figure 5. Panels (a) and (b) in Figures 5 show the full-disk images of the Sun before and after correction, respectively, while panels (c) and (d) provide comparisons of subareas (corresponding to the red-boxed regions), revealing enhanced clarity and effective noise reduction in the corrected images. Figure 5(e) displays the *psf* estimated by SPIBOA, which closely matches the distinctive 'horseshoe' pattern observed in the SDI images, indicating a reasonable and accurate estimation. Figure 5(f) shows the power spectral density profiles of the images in panels (c) and (d), which demonstrate a significant increase in mid- to high-frequency energy in the corrected image compared to the observed image, quantitatively validating the improvements in clarity and contrast.

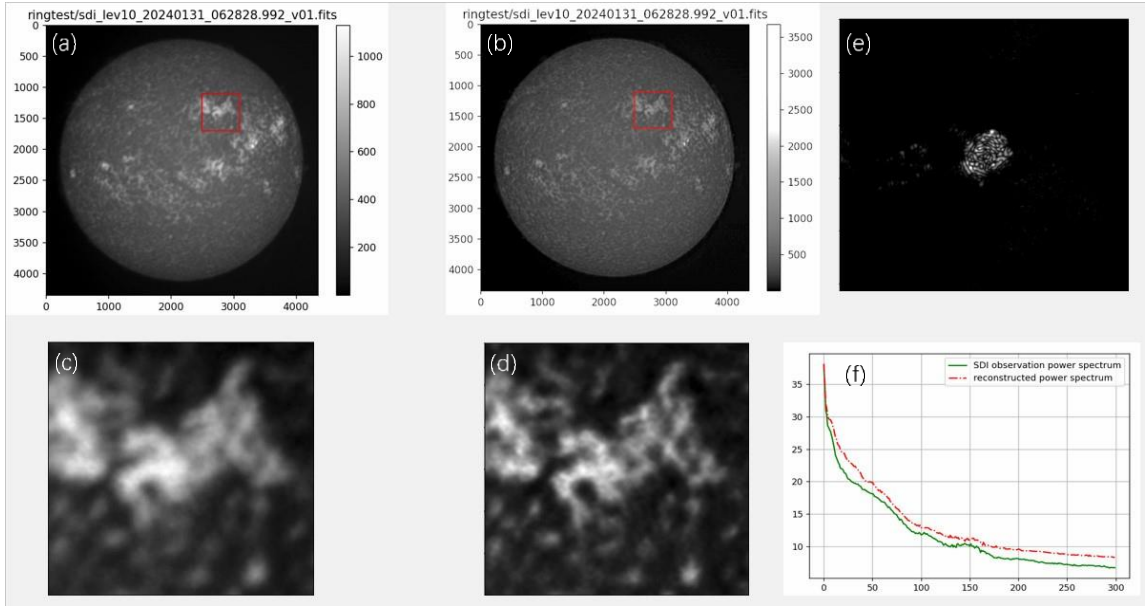

Fig.5: Deconvolution with PSF obtained by SPIBOA and power spectrum. (a) SDI observational image; (b) deconvolution image; images in panels (c) and (d) correspond to the sub-region denoted by the red box in panels (a) and (b), respectively; (e) The *psf* obtained from SPIBOA; (f) The power spectral density profiles of image in panel (c) (green) and panel (d) (red); the abscissa in (f) stands for normalized frequency while the ordinate for power density.

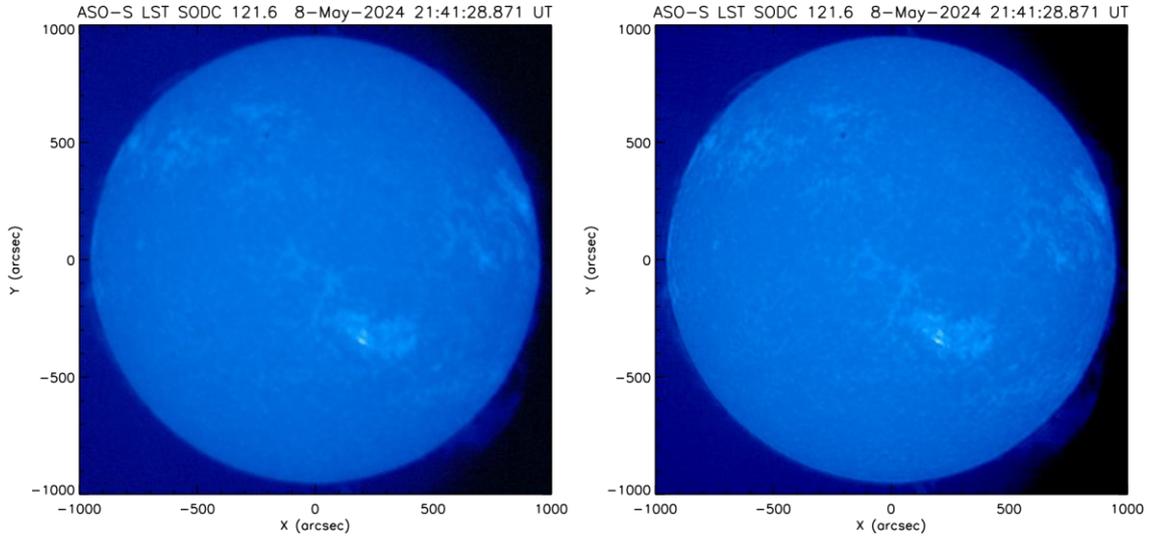

Fig.6: Observed SDI image (left) and reconstructed one (right) taken at 21:41:28 UT on 2024 May 8.

To quantitatively assess the effect of image correction, we employed a series of flare images with varying intensities to measure the resolution improvement after correction. We use the Spatial Image Resolution Assessment by Fourier Analysis (SIRAF) technique (Brostrøm & Mølhave 2022) to conduct the resolution evaluation of SDI images. Our quantitative analysis indicates an enhancement in spatial resolution by a factor of more than three after correction, as outlined in Table 1. The table provides basic information of four flares, including the estimated spatial resolution before and after correction, as well as the factor of improvement in spatial resolution.

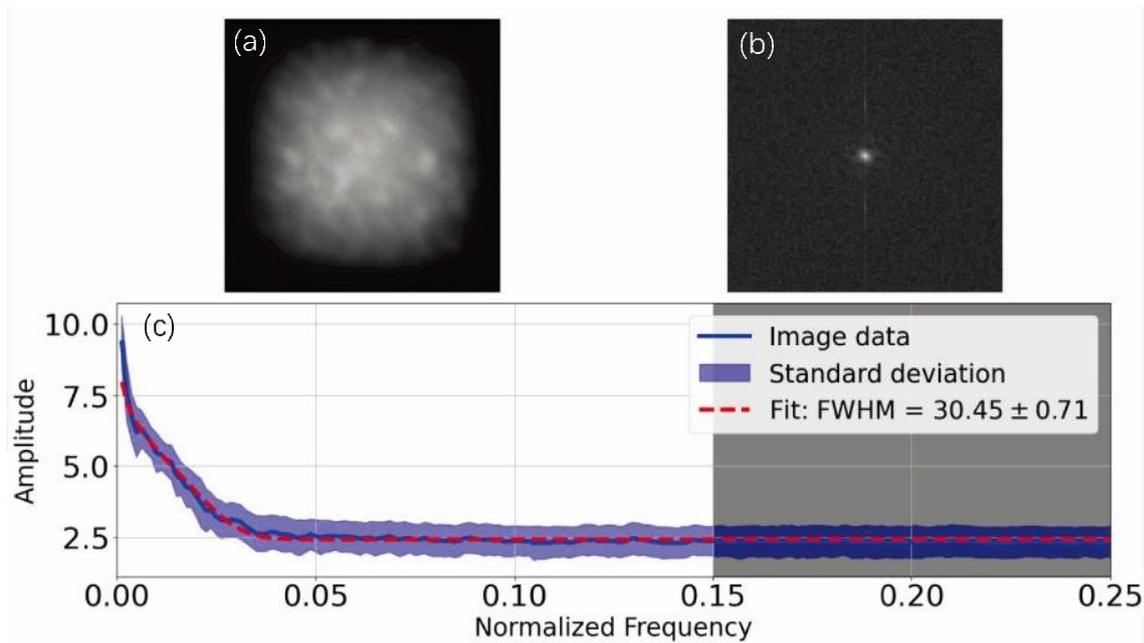

Fig.7: Spatial resolution estimate with SIRAF of SDI image obtained on 2024 May 8. (a) selected region of SDI image filtered using a Hanning window; (b) FFT transformed image in the frequency domain; (c) results of the fitting of the PSF function: its abscissa and ordinate indicate the normalized frequency (in unit length of the reciprocal of image size) and amplitude, respectively. The blue line represents the amplitude spectrum, and the red dashed line represents the fitting curve.

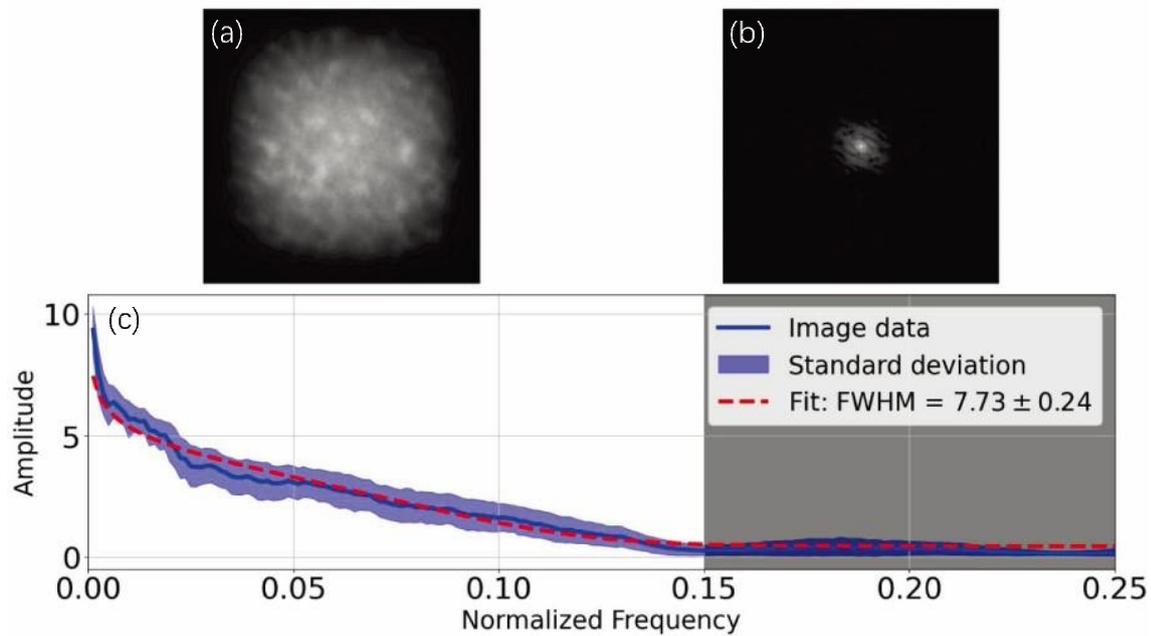

Fig.8: Same as Figure 7 but for the corresponding reconstructed image.

Figures 6 to 8 present resolution comparisons of SDI images before and after correction for the GOES X 1.0 flare of 2024 May 8 in Table 1. Figure 6 displays the full-disk images before and after correction. In Figures 7 and 8, panel (a) shows the windowed images of 800 by 800 pixels, starting at [1600, 1600] in image coordinates, which are used for resolution estimations, and panel (b) the Fourier spectra of these windowed images, whereas panel (c) displays the corresponding spectral amplitude profiles. Following the SIRAF guidelines, a Hanning window is applied to the windowed images to effectively mitigate spectral leakage.

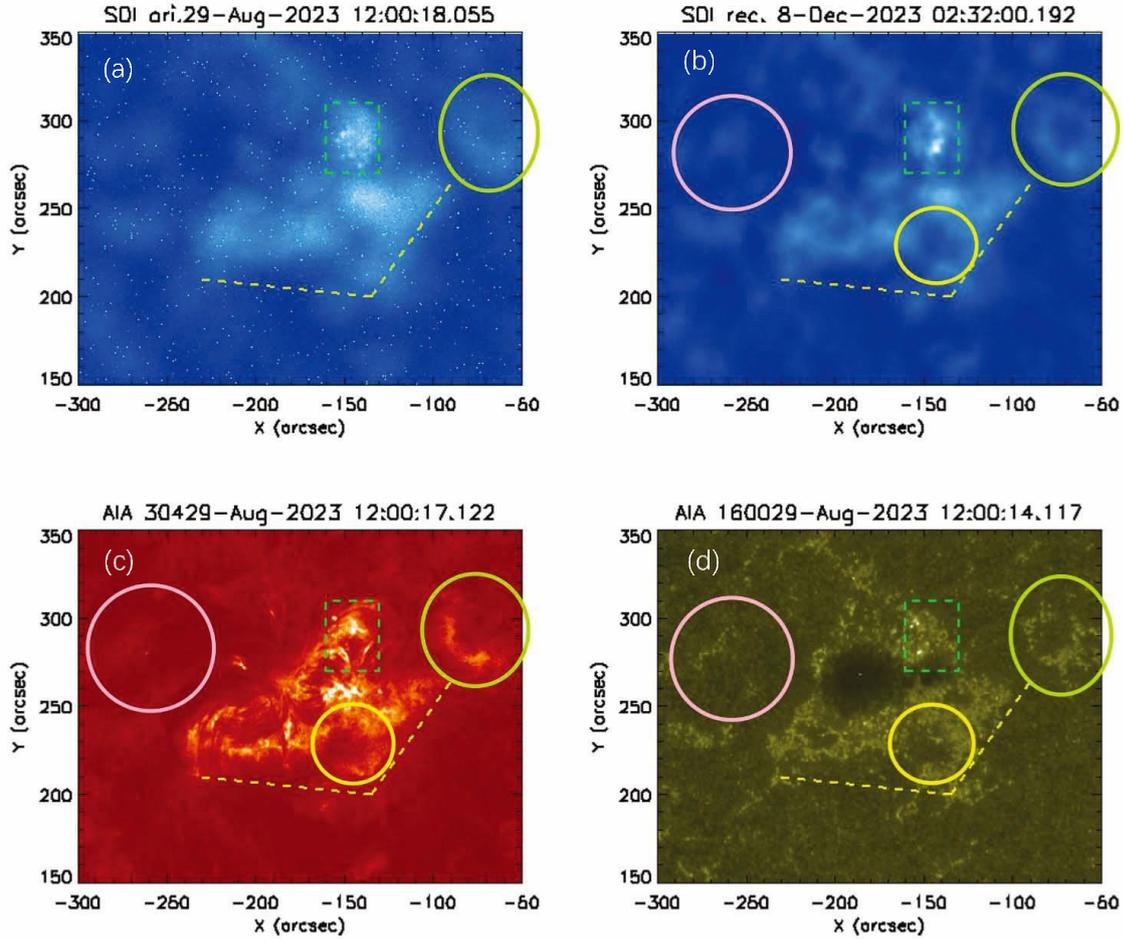

Fig.9: Comparison of (a) original and (b) reconstructed SDI images with (c) AIA 304 A and (d) 1600˚A˚ images.

The corrected images, consistent with the results shown in Figure 5, demonstrate a significant improvement in clarity and contrast, which is due to the fact that the Fourier spectrum of the images before correction was concentrated in the low frequencies, while the mid- to high-frequency energy is noticeably increased after correction. This change can be easily seen by comparing panels (b) in Figures 7 and 8. The spectral amplitude profiles presented in panels (c) of Figures 7 and 8 reveal that the normalized cutoff frequency, which was below 0.05 before correction, rises to approximately 0.15 after correction. The substantial increase quantitatively confirms a more than threefold improvement in spatial resolution.

Figure 9 compares the corrected SDI images with AIA 304 Å and AIA 1600° Å images. The structural° features in the areas marked by squares and circles demonstrate that the spatial resolution of the corrected images has significantly improved. Moreover, these features are well consistent with those in the AIA 304 Å and AIA 1600° Å images, further validating the effectiveness and rationale of the SPIBOA method.

Table 1: Comparison of Spatial Resolution before and after Correction

| Observation Date | Time (UT) | Flare class | Resolution before(″) | Resolution after(″) | Improved by (×) |
|---|---|---|---|---|---|
| 2023 May 9 | 03:58:15 | M6.5 | 12.48 | 4.64 | 2.69 |
| 2023 November 24 | 09:33:34 | M1.2 | 14.05 | 4.27 | 3.29 |
| 2024 February 22 | 22:35:43 | X6.4 | 14.83 | 3.90 | 3.80 |
| 2024 May 08 | 21:41:28 | X1.0 | 15.23 | 3.87 | 3.94 |
| Average |  |  |  |  | 3.43 |

## 5  DISCUSSION

As shown above, the SPIBOA procedure significantly improves the quality of SDI images. In this section, we discuss some important factors, tailored strategies, essential considerations, and precautions for the practical application of the procedure.

### 5.1  The essential elements for the implementation of SPIBOA

- **Prior Constraints:** Utilizing the design and manufacturing parameters of the SDI optical system as fundamental constraints, we determine the equivalent aperture of the imaging system, and then deduce the *psf* of the optical system by refining the wavefront aberration corresponding to this aperture. This approach promotes the efficiency of the optimization procedure and also guarantees that the results more conform to the inherent physical constraints of the optical system. More specifically, the SDI has an aperture of 68 mm and works in the 121.6 nm waveband, corresponding to a diffraction limit of $0.183''$. The designed pixel resolution is approximately $0.5''$ with an FOV of $40'$ and an image dimension of $4608 \times 4608$ pixels. Such a configuration indicates that the digital imaging process is undersampled. Therefore, to accurately estimate the wavefront aberration across the entire aperture, it is imperative to magnify the observed image by a factor of 2.73, equivalent to 0.5 divided by 0.183.

- **Initial Value Selection**: The choice of initial values is crucial for optimization problems. Taking into account the structural similarity and comparable pixel resolution between SDI and AIA 304 Å images° (AIA 304 Å image has a pixel resolution of $0.6°''$, whereas the designed resolution of SDI is $0.5''$), together with that they both capture full-disk images of the Sun, we select AIA 304 Å images that are° temporally close to the SDI observations as the initial values for $I_{ref}$. We subsequently utilize the prowess of CNNs to generate ideal SDI images under optimal imaging conditions. The advantage of this method is its ability to exploit the sophisticated nonlinear fitting capabilities of DNNs, thereby minimizing any potential nonlinear discrepancies between AIA 304 Å and actual SDI images. The alignment is achieved° through linear constraints, which enables a more accurate estimation of the *psf*. During the process, we utilize data obtained from ground-based tests as the initial value of the *wavefront*.

- **Data Samples:** An extensive and diverse dataset is indispensable for training deep learning models. Over the past two years, SDI has accumulated a vast cube of observational data, from which we carefully curated a balanced selection that encompasses data from various solar features: the solar limb, quiet regions, active regions and flare regions ranging from C to X class. Each of these categories is paired with corresponding AIA 304 A images. The balanced training dataset is crucial for optimizing our° model, which has a ratio of 1:1:1:2 for solar limb, quiet regions, active regions and flare regions.

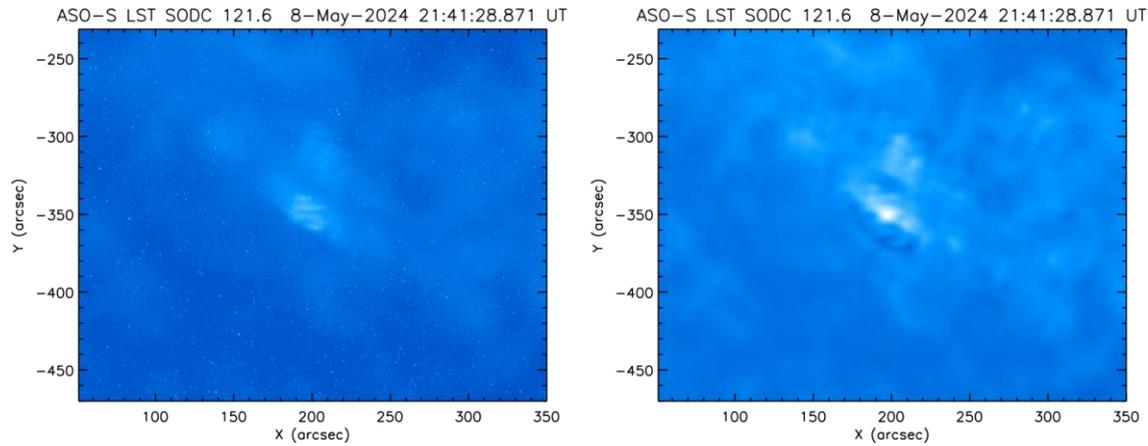

Fig.10: Cutout of observed (left panel) and reconstructed (right panel) SDI image of 2024 May 8. The reconstructed image shows the ring artifacts around the brightening area.

- **Preprocessing:** For the purpose of improving the accuracy of the optimization process and accelerating the training phase, it is necessary to coalign the observed SDI images with the AIA 304 A images within° the same FOV. Additionally, it is important to mitigate the adverse effects of cosmic rays and thermal noise, which can interfere with the image quality and accuracy of the optimization.

## 5.2 Precautions in the Application of SPIBOA

- Image correction is fundamentally a deconvolution process in linear system theory, equivalent to inverse filtering in the frequency domain. However, the so-called 'ringing' artifacts are often introduced due to model inaccuracies and information loss. These artifacts come from the loss of high-frequency image information during degradation, which is particularly noticeable in high-frequency image areas. Severe loss of image information results in steep changes in the corresponding inverse filter, which leads to oscillations of the resulting image in areas with sharp intensity change. Notably, ringing is especially prominent in regions of SDI image with abrupt intensity changes, such as rapid brightening or intense flares. Figure 10 compares the flare region of the observed SDI image on 2024 May 8 with the corrected image. The corrected image (right panel) exhibits typical ringing artifacts, characterized by darker areas with intensity discontinuities within high contrast regions.
- The SPIBOA training program incorporates a wide variety of image samples from a broad temporal range and diverse locations on the Sun into a cohesive training structure, so as to produce a *psf* that effectively captures the average impact of linear distortions across the entire FOV. However, the current version of SPIBOA does not address the temporal variations or spatial inconsistencies that are inherent in SDI imaging system, which may also introduce inaccuracy. This likely explains the varying improvement factors in spatial resolution after applying SPIBOA corrections to the flares listed in Table 1, which occurred at different locations on the Sun. Additionally, SPIBOA currently lacks the ability to provide a precise, quantitative assessment of these potential errors.

## 6  CONCLUSION

This paper concentrates on correcting LST/SDI observational data to enhance image quality, with the specific goals of estimating the *psf* of the SDI imaging system and performing deconvolution corrections on observed SDI images, despite the unknown real SDI image and optical system function. By leveraging a large sample of observational data and employing deep learning techniques, we introduce a bivariate optimization-based imaging model estimation algorithm called SPIBOA.

The outcomes of our research are encouraging and demonstrate a substantial enhancement in the spatial resolution of corrected images. Analysis of sample data has shown an average resolution improvement by over three times, accompanied by a notable decrease in noise levels. We have successfully developed and implemented the SDI data correction software package based on the SPIBOA algorithm, which is currently used to process routine SDI observational data. Furthermore, it incorporates parallel algorithms, which enables efficient and prompt processing of observational data on a single computing node.

We plan to keep our efforts in the future to conduct comprehensive long-term evaluations and enhancements related to the topics discussed in Section 5.2. Additionally, we will be attentive to addressing any emerging challenges. This ongoing commitment will ensure strong support for the scientific utilization of SDI observational data, promoting scientific outputs and advancing reliability in our research endeavors.

The successful deployment of the SDI data correction software package makes it feasible to extend this technique to the other two instruments of the LST payload, namely the WST and SCI, when required. It also pioneers innovative methods for identifying and rectifying issues within the optical systems of astronomical telescopes. Furthermore, it may pave the way for improvements and refinements in astrophysical data and research.

Acknowledgements This work was supported by the National Natural Science Foundation of China (NSFC) under grant No. 12233012, the Strategic Priority Research Program of the Chinese Academy of Sciences, Grant No. XDB0560102, and the National Key R&D Program of China 2022YFF0503003 (2022YFF0503000).

## OCRID iDs

Hui Li  https://orcid.org/0000-0003-1078-3021
Sizhong Zou  https://orcid.org/0000-0001-7607-2594
Kaifan Ji  https://orcid.org/0000-0001-8950-3875
Zhenyu Jin  https://orcid.org/0000-0001-7575-5449
Jianhui Shan  https://orcid.org/0009-0001-4778-5162
Jingwei Li  https://orcid.org/0009-0007-7657-1706
Guanglu Shi  https://orcid.org/0000-0001-7397-455X
Yu Huang  https://orcid.org/0000-0002-0937-7221
Li Feng  https://orcid.org/0000-0003-4655-6939
Jianchao Xue  https://orcid.org/0000-0003-4829-9067
Qiao Li  https://orcid.org/0000-0001-7540-9335
Dechao Song  https://orcid.org/0000-0003-0057-6766
Ying Li  https://orcid.org/0000-0002-8258-4892